# Measuring Innovation Patterns in Iran and Neighboring Countries: A Time Series Similarity Approach Using STL and Dynamic Time Warping

Mahdi Goldani[1]

**Abstract**

Innovation is becoming ever more pivotal to national development strategies but measuring and comparing innovation performance across nations is still a methodological challenges. This research devises a new time-series similarity method that integrates Seasonal-Trend decomposition (STL) with Fast Dynamic Time Warping (DTW) to examine Iran's innovation trends by comparison with its regional peers. Owing to data availability constraints of Global Innovation Index data (2011–2024), research and development (R&D) spending as a proportion of GDP is used as a proxy with its limitations clearly noted. Based on World Bank indicators (2000–2023) and an Autoencoder-based imputation technique for missing values, the research compares cross-country similarities and determines theme domains best aligned with Iran's innovation path. Findings indicate that poverty and health metrics manifest the strongest statistical similarity with R&D spending in Iran, while Saudi Arabia, Oman, and Kuwait show the most similar cross-country proximity. Implications are that Iranian innovation is more intrinsically connected with social development dynamics rather than conventional economic or infrastructure drivers, with region-specific implications for STI policy.

**Introduction**

Innovation has become a key policy priority for governments, aimed at stimulating economic growth and reinforcing institutional capacity (Niazi, 2025). As a cornerstone of development, innovation enhances competitiveness at the institutional, organizational, and national scales (Choi and Zo, 2019). In this regard, science, technology, and innovation (STI) policies represent key governmental strategies aimed at strengthening innovation systems and stimulating innovation activities (Meissner and Kergroach, 2021). Innovation has become a core component of the Sustainable Development Goals (specifically SDG9), which urges nations to "develop resilient infrastructure, advance inclusive and sustainable industrialization, and strengthen innovation." Central to this process is investment in research and development (R&D). For example, SDG9.5 emphasizes the need for countries to significantly increase both public and private R&D expenditures. At the regional level, the highest R&D intensity is observed in the OECD and East Asia, while globally, China and India have emerged as leading hubs of innovation over the past decade (Omar, 2019).

[1] ASSOCIATE PROFESSOR OF ECONOMICS, HAKIM SABZEVARI UNIVERSITY, IRAN EMAIL: M.GOLDANI@HSU.AC.IR

The notion that innovation lies at the core of economic and social development has gained wide recognition. In recent years, numerous low- and middle-income countries have incorporated innovation policies into their national development frameworks, while multilateral organizations have placed innovation programs at the center of policymaking (UNESCO, 2021). This trend builds upon the growing acknowledgment of the role of innovation and innovation capabilities in developing economies an idea that can be traced back to scholarly debates of the 1970s and 1980s (Lema et al, 2021). Table 1 represents the innovation themes in developing countries.

Table1. Innovation Themes in Developing Countries

| THEME / DIMENSION | UPPER-MIDDLE-INCOME COUNTRIES (UMICS) | LOW / LOWER-MIDDLE-INCOME COUNTRIES (LICS/LMICS) | REFERENCES |
|---|---|---|---|
| MANAGEMENT & BUSINESS | Leadership & Knowledge Management: Focus on organizational performance and team creativity (China-driven). | Leadership & Knowledge Management: Team-level innovation in LMIC firms (India/Africa). | Fong et al. (2018) |
| GREEN & ENVIRONMENTAL INNOVATION | Green Industrialization: Eco-innovation, resource efficiency in UMICs. Climate Change & Emissions: Energy efficiency, carbon policy (China strong). | Low-Carbon Development: Tech transfer & sustainability in LICs (Africa, India). | Jin et al. (2019); Shen et al. (2019) |
| DIGITAL TRANSFORMATION | Digital Technologies: E-government, mobile adoption, ICT diffusion. | Limited Digital Adoption in LICs (education & rural tech). | Deng et al. (2010) |
| GLOBAL INTEGRATION & FIRMS | Global Supply Chains: Role of domestic firms in MNCs. | Catching-Up: Capability-building in LMICs. | Story et al. (2015) |
| R&D & KNOWLEDGE SYSTEMS | R&D Networks: Collaboration, commercialization (declining). | Macro-Trends: Trade, diffusion, growth dynamics. | Foreman et al. (2018) |
| HEALTH & SOCIAL SYSTEMS | Health & Education: Service delivery, health innovation. | Health Systems: Governance, rural health inclusion. | Roberton et al. (2015) |
| GOVERNANCE & SOCIETY | Socio-Political Transitions: Institutional reforms (declining). | Social Innovation: Grassroots innovation, community resilience. | Prasad (2016) |
| AGRICULTURE & LIVELIHOODS | Limited role (industrial focus). | Agriculture & Rural Livelihoods: Farm innovation, food security. | Franke et al. (2019). |

According to Table 1, the dimensions of innovation in upper-middle-income countries and low- and lower-middle-income countries vary considerably. Innovation in UMICs is strongly associated with industrial upgrading, green transitions, digitalization, and global competitiveness, while in LICs/LMICs it is more closely tied to basic needs—such as agriculture, health, and social inclusion—often shaped by donor support and external assistance. Analyzing how different countries refine their innovation strategies provides valuable lessons for designing effective plans with minimal costs.

Building on this comparative insight, a more comprehensive assessment of Iran's innovation performance also requires moving beyond domestic data and situating the analysis within a regional framework. Comparing Iran with its neighboring countries in light of their shared geopolitical context offers a clearer understanding of whether the trajectory of Iran's scientific and innovation development is converging with, or diverging from, regional benchmarks. Such a perspective not only highlights whether Iran's knowledge-based and technological sectors are keeping pace with regional trends but also identifies the potential risks of falling behind its peers in the competitive landscape of innovation. In fact, observing the trend of innovation in countries such as Turkey, the United Arab Emirates, Saudi Arabia, Qatar, and Azerbaijan (Figure 1) provides a clearer picture of Iran's position on the regional innovation map. Based on the Global Innovation Index between 2011 and 2024, Iran ranks lower than some of its neighbors, including the UAE, Qatar, Turkey, and Saudi Arabia, and its innovation growth in recent years has been either slow or declining. The Gulf countries, particularly the UAE and Qatar, have secured top positions through stable performance and strong investment in technology. Turkey and Saudi Arabia have also followed an upward trajectory, maintaining relatively high rankings. In contrast, countries such as Pakistan, Tajikistan, and Oman exhibit lower innovation scores and greater fluctuations. Overall, Iran holds a mid-level position, but to compete with the region's leading countries, it needs to strengthen its innovation infrastructure. Such a comparison not only highlights the gap between Iran and its regional rivals but also reveals the strategic opportunities and threats facing the country in shaping its science and technology policies. This study seeks to address this gap by conducting a comprehensive cross-country comparison of innovation dynamics between Iran and its neighboring countries.

Figure 1. Comparison of the Global Innovation Index in Iran and neighboring countries

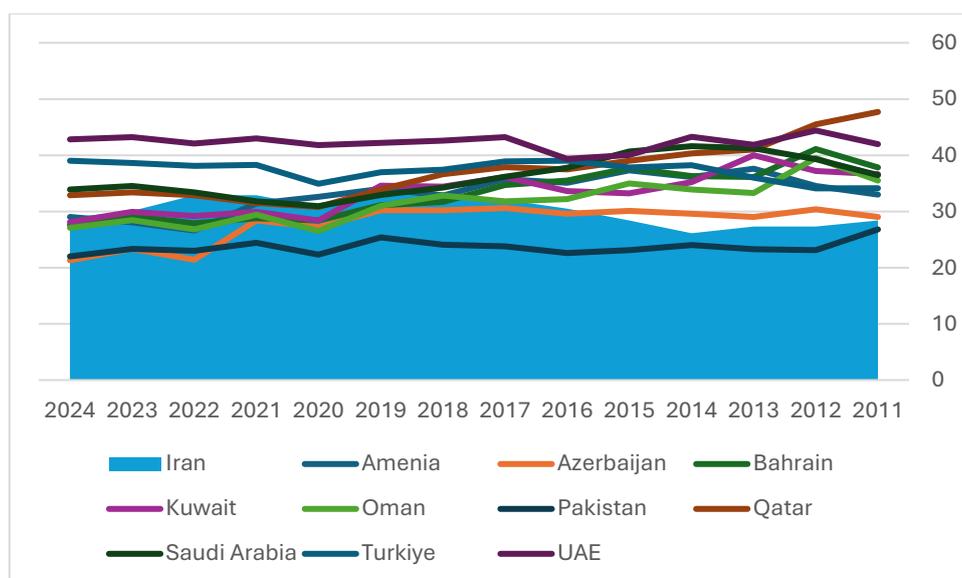

Since preliminary examinations revealed that data for the innovation index were only available from 2011 to 2024, the restricted timeframe risked limiting the robustness of the analysis and potentially leading to misleading conclusions. To address this constraint and improve both the reliability and analytical depth of the study, "research and development expenditure as a percentage of GDP" was adopted as a proxy indicator for the innovation index (Hollanders & Es-Sadki, 2022). Although research and development (R&D) expenditure is among the most popular indicators of innovation capacity, recent literature stresses its limitations in isolation. Ackrill and Çetin (2025) demonstrate that among newly industrialized economies, patents have a more significant long-run impact on high-technology exports than R&D expenditure, highlighting the danger of conflating investment with outcomes. Along the same lines, Ersin, Kirikkaleli, and Ozun (2022) illustrate the non-linear and complex impacts of R&D expenditure, patent activity, and high-technology exports on economic growth, arguing that these variables interact in threshold-dependent rather than linear fashions. Taalbi (2022) warns further that patents are also imperfect proxies, as only a portion of real-world innovations are patented, and their coverage varies dramatically between industries and nations. In an effort to capture the quality dimension of innovation measurement, Ponta (2021) introduces the Innovation Patent Index, adjusting raw patent counts to better reflect the depth and quality of inventive activity. At a policy level, the Organisation for Economic Co-operation and Development (OECD, 2023) has developed new innovation metrics—including patent families assigned and licensing agreements—to enhance coverage of environmental and green innovation, especially in fields associated with climate technology and sustainability. Augmenting these insights, Duarte and Carvalho (2025) undertake a systematic bibliometric analysis of Global Innovation Index (GII) research, demonstrating increased concentration on configurational approaches, efficiency analysis, and policy relevance in innovation research. Cumulatively, this work supports the requirement to acknowledge the multi-dimensionality of innovation measurement and vindicates cautious use of the R&D expenditure proxy indicator, best interpreted against the context of broader socio-economic comparisons.

The purpose of this study is to determine which variables exhibit behavioral patterns that align with the trajectory of the innovation index in Iran and to contrast these patterns with those observed in neighboring and regional countries. Based on the proxy employed in this study, the research questions are formulated as follows:

- RQ1: Which socio-economic variables exhibit temporal patterns most similar to Iran's R&D expenditure?
- RQ2: Which neighboring countries display innovation trajectories most aligned with Iran's?

**Methodology**

**Data collection**

The selection of precise and comprehensive data plays a decisive role in the accuracy and credibility of time series–based analyses. As emphasized in the methodological literature, even the most advanced algorithms cannot compensate for deficiencies arising from poor-quality data. Therefore, in this study, with the aim of achieving temporal comprehensiveness and indicator diversity, data from the World Bank were utilized. This database, with long-term temporal coverage from 1960 to the present and including more than 1,500 economic, social, and environmental indicators, is considered one of the most reliable sources of information for developmental analyses at the international level. In the present study, data related to Iran and its neighboring countries (Azerbaijan, Armenia, Turkey, Turkmenistan, Pakistan, Iraq, Saudi Arabia, the United Arab Emirates, Oman, Kuwait, and Qatar) were extracted and used for the period from 2000 to 2023. Among Iran's neighboring countries, Bahrain was excluded from the analysis due to insufficient data availability, as records were reported for only one year between 2000 and 2023. Each indicator reported by the World Bank is classified under a specific topic, as shown in Table 1. However, to facilitate the interpretation of the results, these categories have been merged into broader groups.

Table 1. classify the topics of world bank development indicators.

|   | group | topic |
|---|---|---|
| 1 | Economic | Economic Policy & Debt: Balance of payments: Capital & financial account |
|   |   | Economic Policy & Debt: Balance of payments: Current account: Transfers |
|   |   | Economic Policy & Debt: Balance of payments: Reserves & other items |
|   |   | Economic Policy & Debt: External debt: Debt outstanding |
|   |   | Economic Policy & Debt: External debt: Debt ratios & other items |
|   |   | Economic Policy & Debt: External debt: Debt service |
|   |   | Economic Policy & Debt: External debt: Net flows |
|   |   | Economic Policy & Debt: National accounts: Adjusted savings & income |
|   |   | Economic Policy & Debt: National accounts: Atlas GNI & GNI per capita |
|   |   | Economic Policy & Debt: National accounts: Growth rates |
|   |   | Economic Policy & Debt: National accounts: Local currency at constant prices: Aggregate indicators |
|   |   | Economic Policy & Debt: National accounts: Local currency at constant prices: Expenditure on GDP |
|   |   | Economic Policy & Debt: National accounts: Local currency at constant prices: Other items |
|   |   | Economic Policy & Debt: National accounts: Local currency at constant prices: Value added |
|   |   | Economic Policy & Debt: National accounts: Local currency at current prices: Aggregate indicators |
|   |   | Economic Policy & Debt: National accounts: Local currency at current prices: Expenditure on GDP |
|   |   | Economic Policy & Debt: National accounts: Local currency at current prices: Value added |
|   |   | Economic Policy & Debt: National accounts: Shares of GDP & other |
|   |   | Economic Policy & Debt: National accounts: US$ at constant 2015 prices: Aggregate indicators |
|   |   | Economic Policy & Debt: National accounts: US$ at constant 2015 prices: Expenditure on GDP |
|   |   | Economic Policy & Debt: National accounts: US$ at constant 2015 prices: Value added |
|   |   | Economic Policy & Debt: National accounts: US$ at current prices: Aggregate indicators |
|   |   | Economic Policy & Debt: National accounts: US$ at current prices: Expenditure on GDP |
|   |   | Economic Policy & Debt: National accounts: US$ at current prices: Value added |
|   |   | Economic Policy & Debt: Official development assistance |
|   |   | Economic Policy & Debt: Purchasing power parity |
| 2 | Education | Education: Efficiency |
|   |   | Education: Inputs |
|   |   | Education: Outcomes |
|   |   | Education: Participation |
| 3 | Employment and Time Use | Employment and Time Use |
| 4 | Environment | Environment: Agricultural production |

| | | | |
|---|---|---|---|
| | | | Environment: Biodiversity & protected areas |
| | | | Environment: Density & urbanization |
| | | | Environment: Emissions |
| | | | Environment: Energy production & use |
| | | | Environment: Freshwater |
| | | | Environment: Land use |
| | | | Environment: Natural resources contribution to GDP |
| 5 | Financial | | Financial Sector: Access |
| | | | Financial Sector: Capital markets |
| | | | Financial Sector: Exchange rates & prices |
| | | | Financial Sector: Interest rates |
| | | | Financial Sector: Monetary holdings (liabilities) |
| 6 | Gender: Public life & decision making | | Gender: Public life & decision making |
| 7 | health | | Health: Disease prevention |
| | | | Health: Health systems |
| | | | Health: Mortality |
| | | | Health: Nutrition |
| | | | Health: Population: Dynamics |
| | | | Health: Reproductive health |
| | | | Health: Risk factors |
| | | | Health: Universal Health Coverage |
| 8 | Infrastructure | | Infrastructure: Communications |
| | | | Infrastructure: Technology |
| | | | Infrastructure: Transportation |
| 9 | poverty | | Poverty: Income distribution |
| | | | Poverty: Poverty rates |
| | | | Poverty: Shared prosperity |
| 10 | Private Sector | | Private Sector & Trade: Exports |
| | | | Private Sector & Trade: Imports |
| | | | Private Sector & Trade: Total merchandise trade |
| | | | Private Sector & Trade: Trade price indices |
| | | | Private Sector & Trade: Travel & tourism |
| 11 | Public Sector | | Public Sector: Conflict & fragility |
| | | | Public Sector: Defense & arms trade |
| | | | Public Sector: Government finance: Deficit & financing |
| | | | Public Sector: Government finance: Expense |
| | | | Public Sector: Government finance: Revenue |
| | | | Public Sector: Policy & institutions |
| 12 | Social Protection | | Social Protection & Labor: Economic activity |
| | | | Social Protection & Labor: Labor force structure |
| | | | Social Protection & Labor: Migration |
| | | | Social Protection & Labor: Unemployment |
| 13 | Trade | | Trade |

The main aim of this article is investigating the innovation pattern and its influenced factors in Iran and compare them Iran's neighboring countries. This comparison is vital for identifying countries whose R&D development policies and trajectories are most similar to those of Iran. These insights may help inform strategies by learning from the experiences of countries with closely aligned patterns. However, preliminary investigations showed that data related to the innovation index were only available for the period from 2011 to 2024. The limited timeframe could reduce the validity of the analysis results and lead to potentially inaccurate conclusions. Therefore, in order to enhance the reliability and depth of the analysis, the indicator "research and development expenditure as a percentage of GDP" was used as an alternative proxy for the innovation index (Hollanders & Es-Sadki, 2022).

**Data Preprocessing**

Given the nature of time series data and the behavioral complexities of macro variables, this study employed an advanced approach for data preprocessing. Specifically, to handle missing values, an Autoencoder-based imputation method was used, which is one of the deep learning techniques applied in data reconstruction. This method can more accurately reconstruct missing values by learning the internal structure of the data. To enhance the performance of the neural networks in this process, the data were standardized using the StandardScaler algorithm prior to being fed into the model, thereby minimizing the impact of differences in variable scales on the model results.

**Similarity Analysis in Time Series Data**

One of the key steps in analyzing time series data is identifying the degree of similarity among the temporal behaviors of different variables. In this study, the main objective of the similarity analysis is to identify development indicators whose patterns over time align with the indicator of research and development expenditure (% of GDP). Identifying dynamically similar patterns among countries can contribute to a deeper understanding of the mechanisms of innovation and help draw lessons from their experiences.

Time series similarity techniques are used to detect dynamic patterns among countries, particularly if the data possess temporal dependencies, nonlinear trends, or potential phase or scale shifts. The techniques generally fall into one of three classes: point-to-point measures, elastic measures, and geometric measures. Point-to-point measures, including Euclidean distance and Pearson correlation, directly compare values at each time step and require perfectly aligned series of identical length. Elastic measures, including Dynamic Time Warping (DTW), LCSS, EDR, ERP, and TWED, are more flexible and stretch sequences with different timing or speed, thus being more resistant to phase shifts, irregular sampling, and noise. Geometric measures, including Fréchet, Hausdorff, and SSPD distances, view time series as trajectories in space and compute similarity from the overall shape and structure of their trajectory. Each of these methods has special strengths in encoding piecewise temporal relations and structural patterns within complex time series data.

DTW is a robust measure for similarity between two series. It has the capability to align sequences so points with similar patterns are mapped even though they occur at different times (figure 1). This elastic warping allows DTW to be used for comparison of time series of different lengths. Additionally, the resulting distance is very robust to outliers (li, 2021). DTW is widely used in theme discovery (Din and Shi, 2017), indexing (Rakthanmanon et al, 2012), gesture recognition (Zadghorban and Nahvi, 2018) etc.

Figure1. Dynamic Time Warping Alignment

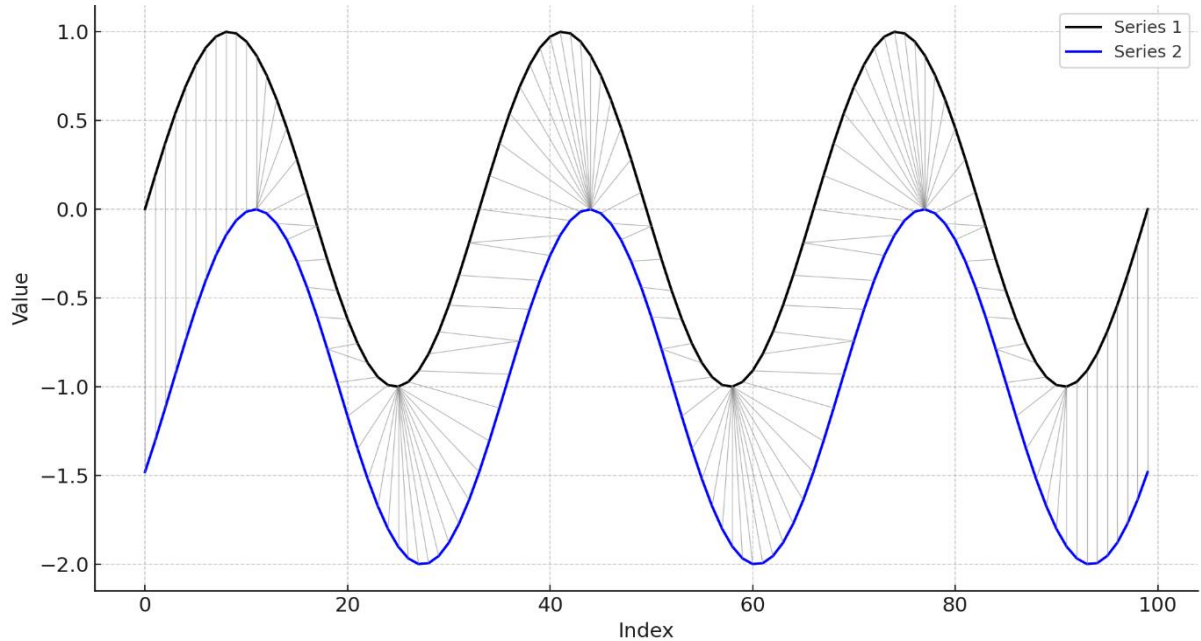

However, DTW's Alignment sequences so points with similar patterns will turn against it, particularly in similarity measurement of long time series. Furthermore, due to over-comparison between time series, regular DTW will lower the overall accuracy of classification, which is known as pathological alignment problem (Zhang et al, 2022).

To overcome the disadvantage of conventional DTW, Fast DTW, by Salvador and Chan, came into being in 2007. The method first applies standard DTW to a low-resolution copy of the time series to derive an initial warping path. The path is then extrapolated and refined at higher resolutions within a bounded window, controlled by a radius parameter. The iterational approach improves alignment accuracy while preserving computation efficiency (Salvador and Chan, 2007).

Time series are comprised of trend, season, and residual components, although all three components do not contribute equally to classification. In some cases, the trend contains more informative information, while in others, the seasonally associated pattern is more important. Traditional DTW-based methods treat all components equally, resulting in more important variations dominating the interesting ones. To handle this, the proposed method decomposes time series and uses Fast DTW to compute similarities between constituents, with the dominant constituent given greater weight, improving classification accuracy.

**Proposed model**

This article by using Zhang et al methodology, first decomposes time series by STL method (Cleveland et al, 1990) and then uses Fast DTW to measure similarities between components, giving more weight to the dominant one, improving classification accuracy. Figure 2 presents the methodology of this research.

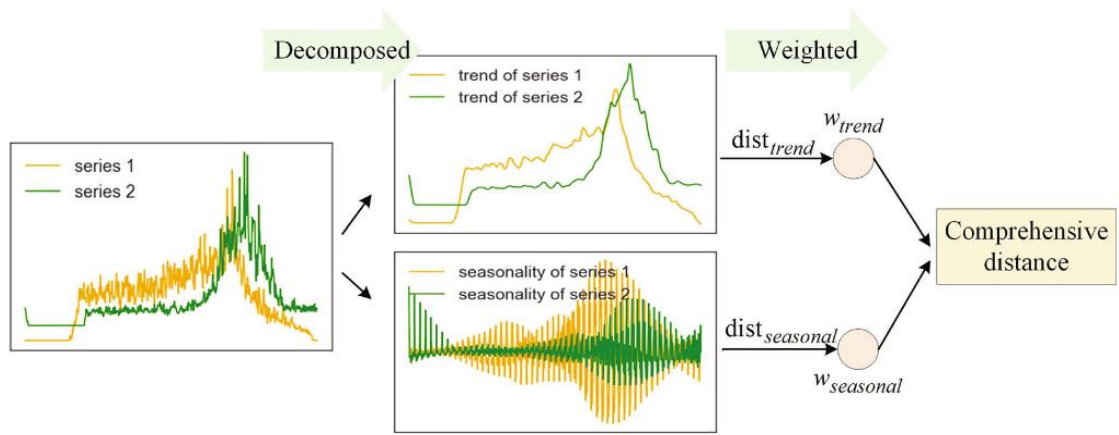

Figure2. Calculation process of the method proposed

Zhang et al (2022)

As discussed earlier, in time series classification, the trend and seasonal components can differ in how much they contribute to distinguishing between series. To reflect this, the method computes two separate distances: one for the trend components ($dist_{\text{trend}}$) and one for the seasonal components ($dist_{\text{seasonal}}$). These distances—between the corresponding components of two series—are then assigned to different weights ($w_{trend}$ and $w_{\text{seasonal}}$) based on their importance. Finally, a weighted combination of the two is calculated to determine the overall similarity between the two-time series, as shown in formula (1).

$$dist(a,b) = dist_{trend}(a,b) * w_{trend} + dist_{\text{seasonal}}(a,b) * w_{\text{seasonal}} \quad (1)$$

Since the method involves decomposing the time series and computing distances twice (once for trend and once for seasonality), the computational cost increases. To maintain efficiency, the Fast DTW algorithm, known for its lower time complexity is used to calculate both dist_trend and dist_seasonal.

**Result**

The aim of this study is to identify variables whose behavior patterns are similar to the innovation index in Iran and to compare these patterns with those observed in Iran's neighboring and regional countries. Understanding the factors that influence innovation in Iran is crucial, as it allows for a comparative analysis with other countries to determine which nations exhibit patterns most akin to that of Iran. Analyzing this similarity provides strategies to foster innovation in Iran by drawing lessons from the successes and challenges faced by peer nations. Initial assessments revealed that data for the innovation index were only accessible for the years 2011 to 2024. This restricted timespan might compromise the accuracy of the analysis and yield less reliable conclusions. To address this limitation and strengthen the robustness and comprehensiveness of the study, the variable "research and development expenditure as a percentage of GDP" was adopted as a substitute proxy for measuring innovation.

For this purpose, combination of Seasonal-Trend decomposition and Dynamic Time Warping (DTW) was selected as the preferred method for identifying indices similar to the research and development expenditure index. This method first decomposed using Seasonal-Trend decomposition via Loess (STL) to separate the long-term trend and seasonal components. The

DTW distance was then calculated separately for the trend and the seasonal components between Iran and each country. To account for the greater importance of the long-term trend in reflecting structural behavior, the distances were combined using weighted aggregation, with 70% weight assigned to the trend distance and 30% to the seasonal distance. The resulting combined distance provided a comprehensive measure of similarity, where lower values indicate stronger similarity to Iran's time series pattern.

As the first step, the research and development expenditure are investigated in selected countries to analyze the pattern of this variable in each country and compare it with Iran. For this reason, the descriptive statistics of research and development expenditure as a percentage of GDP in selected countries are presented in Table 1. The mean of research and development expenditure as a percentage of GDP in Iran is 0.538441, and it shows Iran's moderate record in research and development compared to its advanced neighbors like the UAE and Türkiye. Despite the fact that the UAE has the highest average of research and development expenditure as a percentage of GDP among the regional countries, the standard deviation of the UAE confirms the many ups and downs of research and development expenditure during the investigated years. Iran's performance in research and development expenditure is more stable than that of the UAE and Türkiye. Qatar's and Saudi Arabia's performance are more similar to Iran's performance.

Table 1. Descriptive Statistics of Research and Development Expenditure as a Percentage of GDP in Selected Countries (2000–2023)

|  | mean | std | min | 25% | 50% | 75% | max |
|---|---|---|---|---|---|---|---|
| *Saudi Arabia* | 0.412151 | 0.330088 | 0.0423 | 0.06233 | 0.480595 | 0.685834 | 0.89068 |
| *UAE* | 0.788979 | 0.412417 | 0.47252 | 0.47252 | 0.506305 | 1.139356 | 1.49469 |
| *Turkmenistan* | 0.112917 | 0.016011 | 0.08 | 0.11 | 0.11 | 0.11 | 0.16 |
| *Türkiye* | 0.879412 | 0.327614 | 0.46532 | 0.561087 | 0.80784 | 1.199972 | 1.40209 |
| *Pakistan* | 0.256006 | 0.11126 | 0.09536 | 0.168225 | 0.219968 | 0.311048 | 0.52343 |
| *Qatar* | 0.523878 | 0.071161 | 0.4786 | 0.4786 | 0.4786 | 0.528175 | 0.68058 |
| *Oman* | 0.17952 | 0.075806 | 0.11962 | 0.11962 | 0.13408 | 0.220722 | 0.36729 |
| *Kuwait* | 0.135647 | 0.081369 | 0.06349 | 0.085765 | 0.101115 | 0.152313 | 0.42705 |
| *Iraq* | 0.041205 | 0.004397 | 0.03266 | 0.037316 | 0.0407 | 0.04554 | 0.04596 |
| *Iran* | 0.538441 | 0.185289 | 0.23601 | 0.374103 | 0.56802 | 0.72665 | 0.78829 |
| *Azerbaijan* | 0.22516 | 0.054963 | 0.1513 | 0.18455 | 0.21031 | 0.230365 | 0.34001 |
| *Armenia* | 0.234299 | 0.035296 | 0.17854 | 0.209462 | 0.23579 | 0.25356 | 0.32109 |

To gain deeper insight into the trends of research and development expenditure as a percentage of GDP in selected countries during 2000–2023, the similarity between these variables was measured to determine which countries' performance is most similar to Iran's performance. Qatar is the country whose behavior in research and development expenditure, based on the trend DTW and combined DTW, is most similar to that of Iran. In terms of seasonal DTW, Armenia, Turkmenistan, and Türkiye exhibit similar behavior to Iran.

Figure 1. Dynamic Time Warping (DTW) Distances Between Iran and Selected Countries for Trend and Seasonal Components of R&D Expenditure

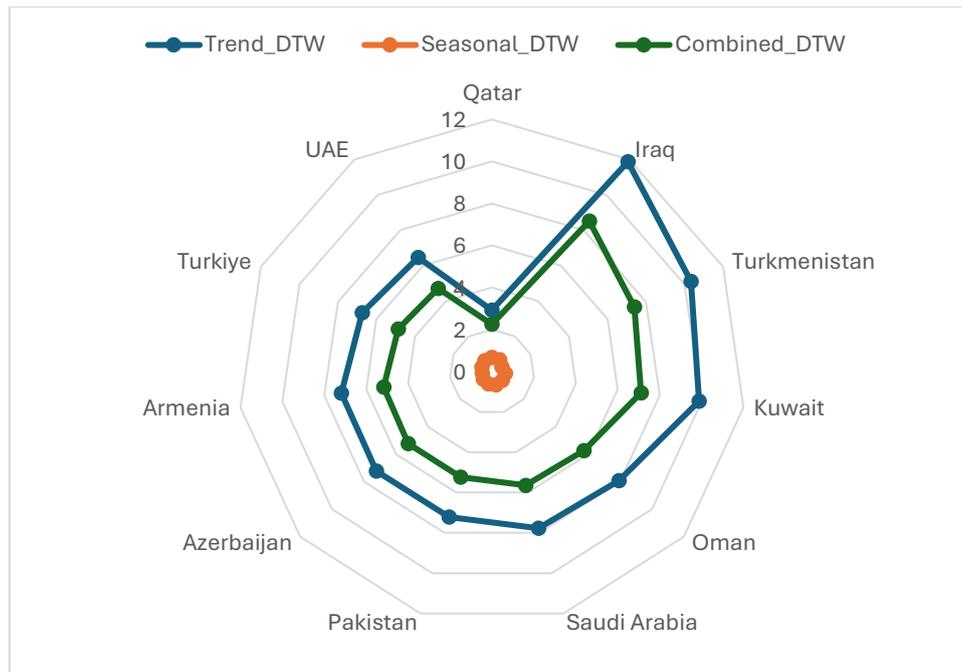

In this section, the world bank development indicators have been ranked according to their similarity to R&D Expenditure in selected countries target value. Figure 2 presents the classification of variables based on World Bank topics in Iran, offering a comprehensive view on their similarity to R&D expenditure. Here, the mean rank for every category is the average rank of variable rank in each category and lower mean ranks are an indication of higher similarity with R&D spending.

The Poverty and health categories have the lowest mean ranks (248.1667 and 370.1987), i.e., the highest similarity to R&D expenditure. Poverty does have, however, a relatively low number of variables, i.e., despite being strongly similar, they are less important overall. The health category comprises nearly 19% of all variables and achieves one of the lowest average ranks—second only to the Poverty group—indicating a high degree of similarity with the research and development expenditure index in Iran. In other words, variables in the health category have had a greater influence on R&D expenditure than those in other categories. Health and poverty variables are statistically most similar to R&D expenditure. However, this does not imply causation, only temporal co-movement. The Economic Policy & Debt category, despite having the highest percentage of variables (24.67%), has the highest mean rank (691.854), indicating weaker similarity overall. Education, Financial Sector, and Infrastructure categories are shown by greater mean ranks, demonstrating a relatively lower similarity with R&D expenditure despite constituting a larger proportion of the dataset.

Overall, this distribution points to the fact that the quantity of variables within a category will not necessarily be directly proportional to the degree of similarity with R&D spending. This insight provides valuable guidance for sector and topic prioritization, most resembling research and development spending.

Fig2. Grouping of Variables Resembling R&D Expenditure in Iran According to Their Similarity Scores

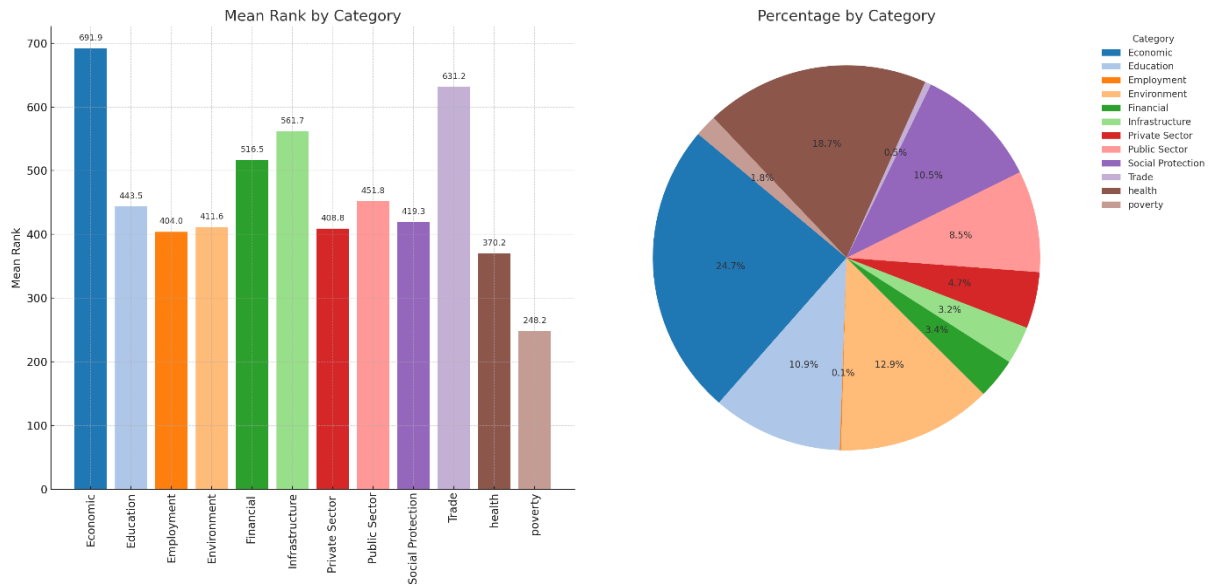

Table 2 presents the similarity of ranking patterns of variables related to research and development (R&D) expenditures between Iran and other countries. In this table, countries with the lowest DTW distances demonstrate a higher similarity in the ranking patterns of factors compared to Iran. Saudi Arabia, Oman, and Kuwait are three neighboring countries whose ranking patterns are more similar to Iran's.

Table 2. Dynamic Time Warping (DTW) Distance Between R&D Variable Rankings in Iran and Other Countries

|    | **COUNTRY**  | **DTW-DISTANCE** |
|----|--------------|------------------|
| 1  | Saudi Arabia | 106906           |
| 2  | Oman         | 123789           |
| 3  | Kuwait       | 128176           |
| 4  | UAE          | 131820           |
| 5  | Turkmenistan | 132759           |
| 6  | Qatar        | 141734           |
| 7  | Türkiye      | 150183           |
| 8  | Iraq         | 167813           |
| 9  | Azerbaijan   | 174252           |
| 10 | Armenia      | 184936           |
| 11 | Pakistan     | 242858           |

**Conclusion**

The primary objective of this article is to examine innovation patterns and their driving factors in Iran and to compare them with those of Iran's neighboring countries. Such a comparison is crucial for identifying nations with R&D policies and development paths most comparable to Iran's, providing opportunities to draw lessons from their experiences. Nonetheless, initial reviews revealed that data on the innovation index were only accessible for the years 2011 to

2024. This restricted timeframe may weaken the robustness of the findings and risk leading to inaccurate interpretations. To strengthen the reliability and depth of the analysis, the study instead employs "research and development expenditure as a percentage of GDP" as a proxy indicator for innovation. To ensure both temporal comprehensiveness and diversity of indicators, this study relied on data provided by the World Bank. This database, which offers extensive coverage from 1960 to the present and contains more than 1,500 economic, social, and environmental indicators, is widely recognized as one of the most credible sources for international development analysis. For the purposes of this research, data pertaining to Iran and its neighboring countries (Azerbaijan, Armenia, Turkey, Turkmenistan, Pakistan, Iraq, Saudi Arabia, the United Arab Emirates, Oman, Kuwait, and Qatar) were collected for the period 2000–2023. Bahrain was excluded from the study due to insufficient records, as data were available for only one year within the selected timeframe. While the World Bank assigns each indicator to a specific topic, in this study the categories were consolidated into broader groups to facilitate clearer interpretation of the findings.

The investigation found that at least within the World Bank development indicators, variables capturing poverty and health showed the closest match to R&D spending in Iran, albeit in relative terms for their share of total indicators. This might suggest that areas more closely linked to social well-being have had more effect on Iran's R&D expenditure than the more general categories of - economic policy, debt, and infrastructure in which the association is relatively weaker. The results demonstrate that the number of variables vetted in a category did not always clarify their importance for R&D trends.

The cross-country case comparison using Dynamic Time Warping (DTW) further shows that Iran possesses the most similar ranking patterns of R&D related variables with Saudi Arabia, Oman and Kuwait. This may offer possibilities for Iran to be informed by the developmental policies and models of innovation mobility in these countries (and any innovations in how to deploy R&D in social economic priority settings) in forming its own plans for R&D strategy.